\documentclass[journal]{IEEEtran}

\usepackage{graphicx}
\usepackage{tabularx}
\usepackage{lipsum}
\usepackage{array}
\usepackage{color}
\usepackage{xcolor}
\usepackage{booktabs}
\usepackage{bm}
\usepackage{textcase}
\usepackage{multirow}
\usepackage{cuted}
\usepackage{mathtools}
\newcommand{\tabincell}[2]{\begin{tabular}{@{}#1@{}}#2\end{tabular}}

\makeatletter
\def\endthebibliography{%
	\def\@noitemerr{\@latex@warning{Empty `thebibliography' environment}}%
	\endlist
}
\makeatother

\begin{document}
\title{Multi-task Learning Based Spoofing-Robust Automatic Speaker Verification System}
\author{Yuanjun~Zhao,~\IEEEmembership{Student~Member,~IEEE,}
	    Roberto~Togneri,~\IEEEmembership{Senior~Member,~IEEE}
        and~Victor~Sreeram,~\IEEEmembership{Senior~Member,~IEEE}
\thanks{Yuanjun Zhao, Roberto Togneri and Victor Sreeram are with the Department of Electrical, Electronic and Computer Engineering, The University of Western Australia, Perth, WA 6009, Australia. }
\thanks{E-mail: yuanjun.zhao@research.uwa.edu.au.}
}

\markboth{Journal of \LaTeX\ Class Files,~Vol.~14, No.~8, August~2015}%
{Shell \MakeLowercase{\textit{et al.}}: Bare Demo of IEEEtran.cls for IEEE Journals}

\maketitle

\begin{abstract}
Spoofing attacks posed by generating artificial speech can severely degrade the performance of a speaker verification system. Recently, many anti-spoofing countermeasures have been proposed for detecting varying types of attacks from synthetic speech to replay presentations. While there are numerous effective defenses reported on standalone anti-spoofing solutions, the integration for speaker verification and spoofing detection systems has obvious benefits. In this paper, we propose a spoofing-robust automatic speaker verification (SR-ASV) system for diverse attacks based on a multi-task learning architecture. This deep learning based model is jointly trained with time-frequency representations from utterances to provide recognition decisions for both tasks simultaneously. Compared with other state-of-the-art systems on the ASVspoof 2017 and 2019 corpora, a substantial improvement of the combined system under different spoofing conditions can be obtained.   
\end{abstract}


\IEEEpeerreviewmaketitle

\section{Introduction}
\label{sec_intro}

\IEEEPARstart{P}{rior} to the consideration of spoofing, speaker/voice recognition systems have been designed and widely used for commercial and forensic applications by identifying and verifying the claimed identity of a speaker \cite{togneri2011overview}. However, for instant and convenient authentications, issues of malicious interference and manipulations on speaker recognition systems are coexisting \cite{zhao2019applications}. The potential for speaker recognition systems to be spoofed is now well-recognized \cite{evans2014speaker,evans2013spoofing,wu2013vulnerability}. Urgent needs are suggested to address spoofing in numerous vulnerability studies. Generally, approaches involved for anti-spoofing concentrate on proposing specific or generalized standalone spoofing detectors. From economical and practical perspectives, designing spoofing-robust systems which integrate the functions of spoofing detection and speaker recognition make sense. While both options are vital in the community, integrated systems can assist to streamline recognition processes, reduce costs and ensure efficiency. 

Until now, most relevant studies only focus on the framework and evaluation of standalone countermeasures. However, there are numbers of reasons that the integration of the spoofing detection and speaker recognition system is important \cite{sahidullah2019introduction}. First, since the recognition system and its corresponding spoofing detector are trained to solve two different tasks, a standard linear fusion on the score level is not appropriate. Second, the performance of a spoofing detection system is critical for the final output decision. An imperfect spoofing detector can increase the false alarm rate by rejecting genuine speakers \cite{sahidullah2016integrated}. Thirdly, in a framework which contains separated spoofing detection and speaker recognition modules, it has not been confirmed whether improvements in standalone countermeasures should improve the overall system as a whole. A perfect anti-spoofing system will fail to protect a recognition system which is not properly calibrated \cite{muckenhirn2017long}. 

\begin{figure}[htbp]
	\centering
	\includegraphics[width=\linewidth]{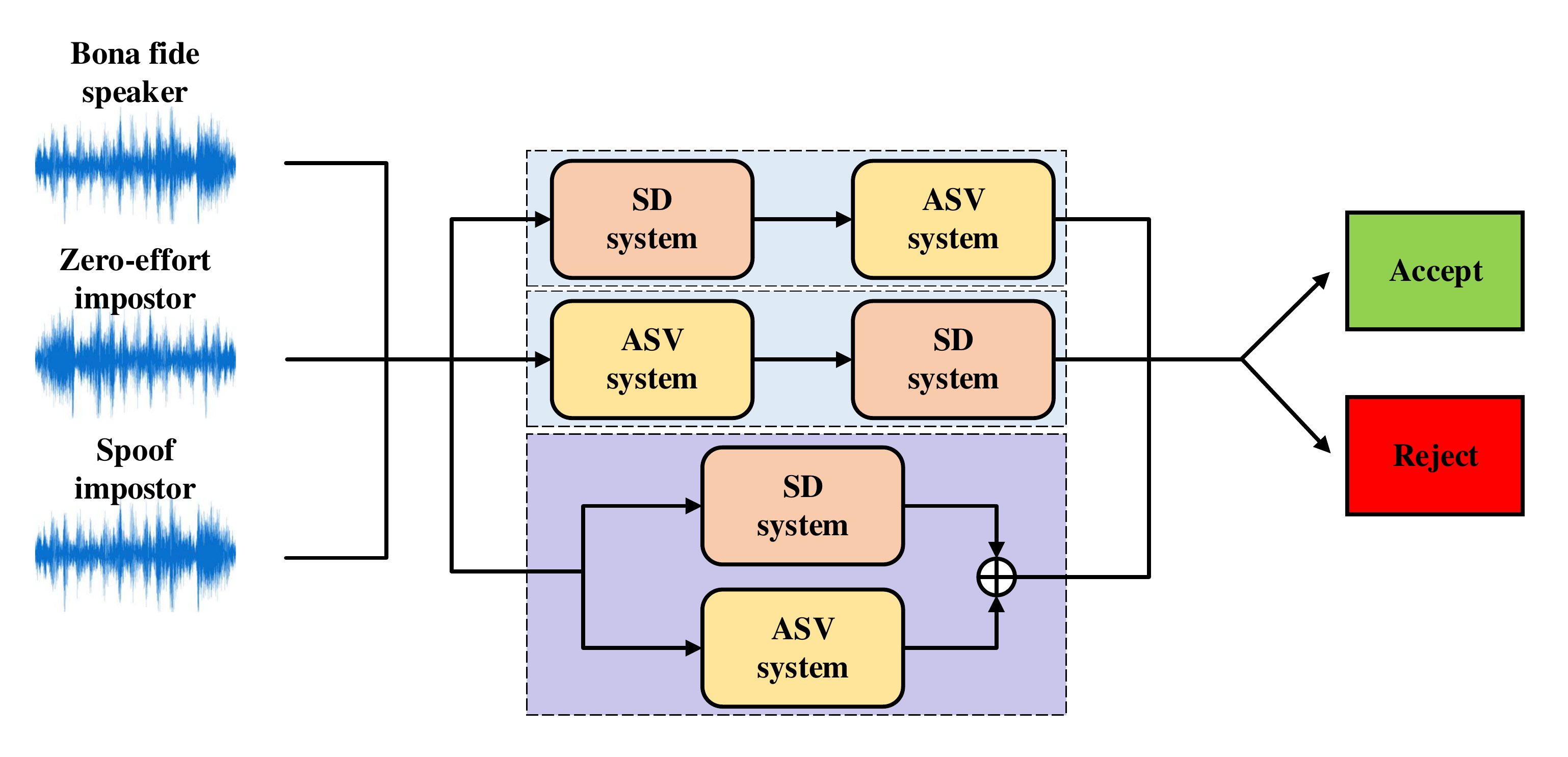}
	\caption{Integrated systems of different combination methods.}
	\label{fig:allsystems_asv}
\end{figure}
The impact of standalone spoofing detection systems is difficult to be gauged unless they are evaluated when integrated with an ASV system \cite{hadid2015biometrics}. For example, as binary classification problems, ASV and spoofing detection systems have a common goal of distinguishing unauthorized access attempts. For ASV systems, zero-effort impostors are rejected. While for spoofing detectors, forged trials from the spoof impostors will be detected. As shown in Fig.\ref{fig:allsystems_asv}, there are two conventional and simple solutions to jointly combine an ASV and a spoofing detector. The first is to cascade these two modules in alternative orders to protect ASV from spoofing attacks \cite{alegre2013spoofing,de2012evaluation}. Obviously, two recognition thresholds are required in the cascaded approach to be applied to each module. The final decision is obtained by comparing the produced scores with two thresholds. Only the trials with scores that are not less than both the thresholds can be accepted. In addition to the cascaded method, another solution is the parallel approach which is also shown in Fig.\ref{fig:allsystems_asv}. A single threshold is used in the parallel approach to compare with fused scores for the final decision. For integrated systems, only Bona fide speech samples can be accepted for correct authentications while samples from both the zero-effort and spoof impostors will be rejected as illegal access. 

Research in the area of spoofing-robust integrated recognition system is still in its relative infancy and greater attention is needed in the future. In \cite{sahidullah2016integrated}, several state-of-the-art spoofing countermeasures were integrated with ASV systems. Selected countermeasures were combined in a cascaded or parallel framework and evaluated with the ASVspoof 2015 corpus. Experimental results indicated that when ASV systems were integrated with a diverse set of countermeasures, the performance can remain robust in the presence of varying attack approaches. A subsequent exploration on the ASVspoof 2017 V2.0 corpus was given in \cite{Todisco2018}. A Gaussian back-end fusion approach was presented to combine the spoofing detector and ASV system. A variety of different features were used to assess the performance of the integrated system. The proposed combination approach was shown to generalize particularly well across independent development and evaluation subsets. In \cite{dhanush2017factor}, a joint modeling approach was introduced to detect spoofing attacks while also performing the speaker verification task. Factor analysis methods were adopted such that the spoof variability subspace and the speaker variability subspace are jointly trained. Experiments were performed using the speaker and spoofing (SAS) database \cite{wu2015sas}. Compared to a baseline system integrated in the conventional method, the proposed approach provided substantial improvements for spoof detection as well as speaker verification. 

Unlike conventional fusion methods, the problem of ASV and spoofing detection integration is that these two systems are designed with different objectives. It has been demonstrated that the performance of a spoofing detector naturally impacts the performance of the ASV system; either the false alarm rate or the false reject rate will be influenced \cite{Kinnunen2018}. With progress in standalone anti-spoofing research continuing, we should also concentrate on integrated spoofing-robust ASV systems which are optimized jointly. Another problem relates to the degraded performance of ASV systems which are expected to be attacked by varying types of spoofed speech. Even if a speaker verification system is integrated with spoofing detection, it is still troublesome to overcome the performance loss caused by diverse attacks. Although integrated spoofing detection and ASV systems have been proposed no system has been designed to handle diverse attacks, that is both logical access (using machine generated spoofed speech) and physical access (using replayed spoofed speech) attacks.  

Different to the previous works, in this paper, we pursue solutions for integrated spoofing-robust ASV (SR-ASV) systems which are aware of the logical and physical access attacks simultaneously. To extend the generalization ability of the model used, sequential residual convolutional blocks with Max-Feature-Map activations (MFM) \cite{wu2018light} are applied. So far, however, there has been limited discussion about this type of versatile anti-spoofing countermeasure which can handle both condition attacks. The work presented here provides the first investigation on how to jointly optimize both the spoofing detection and ASV tasks for diverse attacks. The proposed integrated system is evaluated on the newly released ASVspoof 2017 Version 2.0 and 2019 corpora and compared with other state-of-the-art systems. Results demonstrate that the proposed SR-ASV system can overwhelm the other state-of-the-art integrated systems for both spoofing conditions. This also indicates the model used is efficient for both speech processing tasks. Detailed discussion and analysis of the experimental results are given in Section~\ref{sec_ERA}. More details of the proposed SR-ASV system can be found in Section~\ref{sec_SRASV}. 

The contributions of this paper are as follows:
\begin{itemize}
	\item In this paper, it is the first time that an integrated spoofing-robust ASV system is proposed with a generalization for both logical and physical condition attacks. By adopting the multi-task learning, the system introduced is optimized jointly to obtain an effective representation based on the combined information from anti-spoofing and speaker verification tasks. The auxiliary relations between these two tasks are utilized in the training process. 
	\item The discriminative information of speakers and artifacts caused by spoofing attacks in acoustic features are crucial for building an effective verification system. To obtain the abilities of reducing spectral variations and modeling spectral correlations in acoustic features, we adopt sequential residual convolutional blocks with MFM activations. These network units are used for the first time in the training for integrated spoofing-robust ASV systems based on the multi-task learning. 
\end{itemize}

The rest of the paper is organized as follows. In section~\ref{sec_liter}, several related works are introduced. The proposed SR-ASV system is introduced in section~\ref{sec_SRASV} and the experiment settings are provided in section~\ref{sec_expsetting}. The experimental results and relevant analysis are given in section~\ref{sec_ERA} followed by our conclusions in section~\ref{sec_end}. 

\section{Related works}
\label{sec_liter}

In this paper, we introduce a novel integrated solution for a spoofing-robust ASV system based on deep learning techniques. Generally, deep neural networks (DNNs) are used for extracting discriminative embeddings for each speaker \cite{snyder2017deep,snyder2016deep} and as part of an end-to-end system for speaker verification \cite{heigold2016end}. Applying a deep learning based architecture to speaker verification is a relatively new endeavor \cite{nassif2019speech}. The proposed system is based on the multi-task learning (MTL) approach, which has been used successfully across all applications of machine learning, from natural language processing \cite{sogaard2016deep} and speech recognition \cite{chen2015multitask} to computer vision \cite{Kendall_2018_CVPR} and acoustic event detection \cite{xia2019multi}. Recently, multi-task learning based architectures have been adopted in biometrics recognition and anti-spoofing (especially for presentation attack detection (PAD) which is also known as replay spoofing). In \cite{chen2018multi}, a multi-task PAD approach was proposed to simultaneously perform iris detection and iris presentation attack detection. A convolution neural network (CNN) used for general object detection was leveraged to build a multi-task learning framework. With this approach, a bounding box defining the spatial location of the iris can be predicted and a presentation attack score denoting the probability can be generated. The MTL framework was also used for joint face recognition and PAD \cite{kambicmulti}. Convolutional layers were applied for feature extraction and two parallel output networks were used for face recognition and classification of PAD.

Due to the effective representation learned by MTL models, this architecture has also been adopted in speaker anti-spoofing. In \cite{li2019anti}, an MTL network was used for improving anti-spoofing performance with a proposed helpful butterfly unit (BU). The authors achieved the evaluation EER of 2.39\% from the best single system on ASVspoof 2019 PA. In \cite{von2020multi}, a siamese neural network (SNN) was used to build an MTL network that can yield improvement with additional reconstruction loss. In addition, multi-task outputs can also be applied to predict spoofing labels and replay configuration labels as in \cite{Yang2019}. The sum of the multi-task outputs in both Bona fide nodes was regarded as the detection score. A similar work was \cite{shim2018replay}, in which an MTL network was used to classify the noise of playback devices, recording environments and recording devices as well as the spoofing detection.

In \cite{Todisco2018} it was the first time to fuse anti-spoofing and speaker verification on the score level by using a cascade manner, while with the MTL network the integration can happen on the model level. Applying MTL models for integrating anti-spoofing and speaker verification is still in the early stages. In \cite{li2020joint}, contrastive loss was used in an MTL framework in order to improve the cascaded decision approach. A modified triplet loss was constructed for extracting embeddings containing information of both speaker identity and spoofing. However, this work only focused on the physical condition attacks and the logical condition attacks were not considered, which can also pose a serious threat to ASV systems. In addition, the deep learning architectures adopted in \cite{li2020joint} were conventional networks in speaker recognition, such as sequential fully connected layers, convolutional layers and time-delay DNNs. Although these networks have been proved as efficient solutions for the ASV task, their effectiveness in spoofing detection is still unknown. 

In this paper, we provide a more effective MTL framework by employing the Max-Feature-Map activation (MFM) \cite{wu2018light} which has been used in high-performing LCNN based spoofing detection systems \cite{lavrentyeva2017audio,lavrentyeva2019stc}. We also design the entire MTL network according to the ResBlock used for ASV in \cite{li2017deep}. With these architectures, the proposed network can be trained to extract speech embeddings of high discrimination and precise representation. The type of embeddings extracted include both the information used for spoofing detection and ASV tasks and will not cause biased evaluation results. It is the first time that residual convolutional blocks and MFM activation layers are used in a multi-task learning framework for jointly training a spoofing-robust speaker verification system. More introductions of the network used in this paper are given in Section~\ref{sec_SRASV}.


The multi-task learning approach is performed with hard parameter sharing of hidden layers in this network. The tasks of spoofing detection and speaker verification are optimized jointly with an auxiliary effect. Hidden layers are shared between these two tasks during the training while several task-specific output layers are cascaded afterwards. This is the most commonly used approach in multi-task learning to greatly reduce the risk of over-fitting. Since the tasks of spoofing detection and speaker verification are learned simultaneously, the model is forced to find a representation that captures both tasks and generalizes better than either single task. After training, the model will be used for extracting embeddings applied with a probabilistic linear discriminant analysis (PLDA) back-end for speaker verification. The decisions for spoofing detection can then be directly achieved by adding a softmax output layer to the network or adopting a GMM back-end classifier.  

\section{Proposed SR-ASV system}
\label{sec_SRASV}

\subsection{Framework of SR-ASV system}

\begin{figure*}[htbp]
	\centering
	\includegraphics[width=11cm,height=10cm]{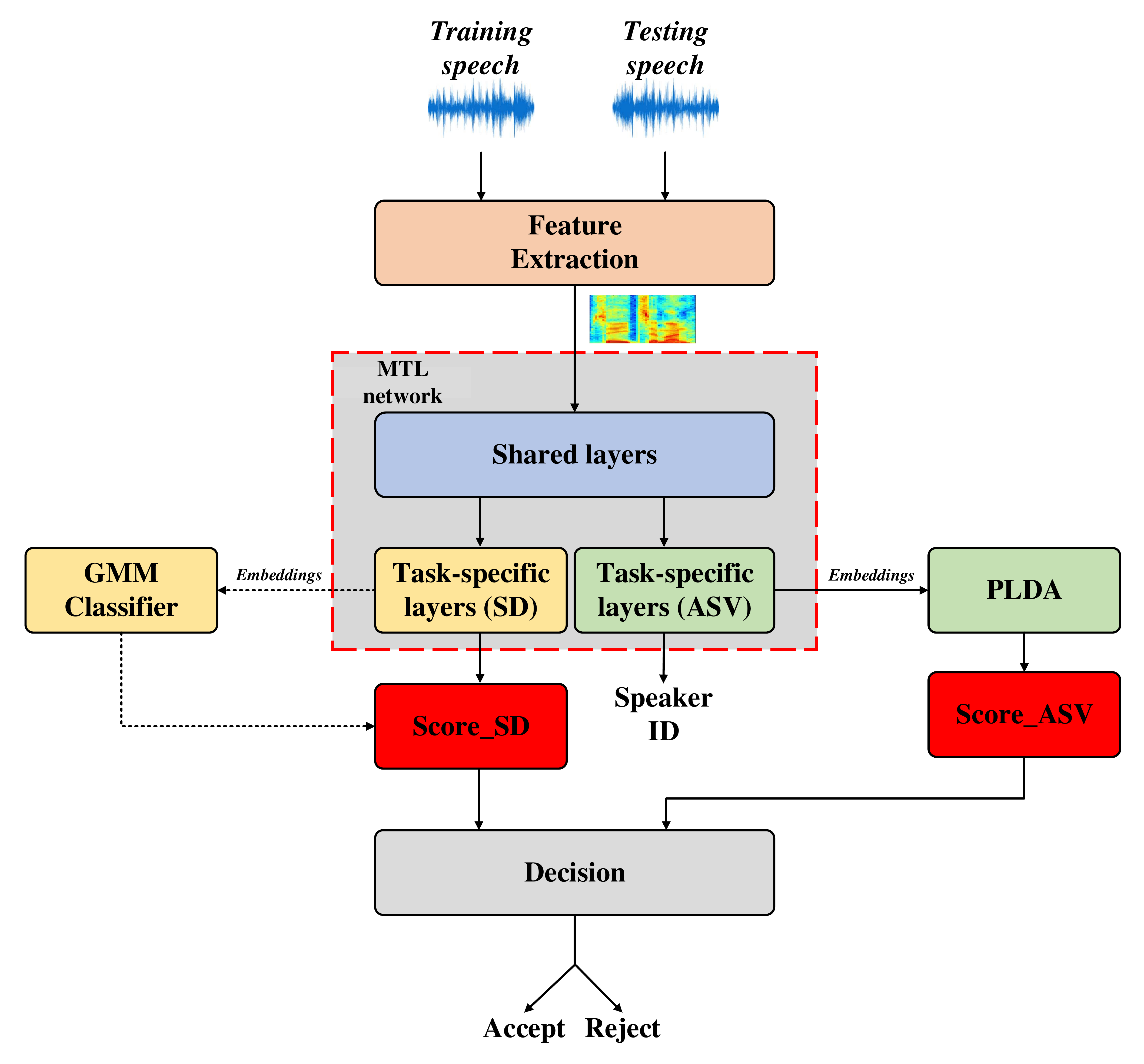}
	\caption{The scheme of the proposed SR-ASV system.}
	\label{fig:MTLsystem}
\end{figure*}

In this section, the detailed framework of the proposed SR-ASV system is introduced, which is depicted in Fig.\ref{fig:MTLsystem}. In the preprocessing phase, the training samples are converted to time-frequency representation based features. Extracted time-frequency representation (TFR) features are used to train the deep learning based embedding extractor with corresponding labels which contain the speaker IDs and detection keys. The MTL network used in this work is designed according to the high-performing LCNN architecture \cite{lavrentyeva2017audio,lavrentyeva2019stc} and the ResBlock employed in the Deep Speaker \cite{li2017deep}.

For the task of ASV, a softmax layer is applied in the training process and the number of nodes equals the number of training speakers. Following several fully connected layers, this output layer can provide discriminative embeddings for each speaker. These representations are centered and length-normalized before modeled by a PLDA back-end. The scores are normalized using the adaptive s-norm \cite{snyder2018x}. To achieve the final decision, we use the t-DCF \cite{Kinnunen2018} as the performance metric which will be introduced later.

Compared to shallow networks, deep networks can learn complex representations from acoustic features and even original time-domain waveforms. However, it is commonly known that deep networks tend to be more difficult to train due to problems like vanishing gradient. To overcome this, stacked residual blocks are used to force the gradient information to be passed to deeper layers in the network. Each ResBlock contains direct links between the shallower layer outputs and the deeper layer inputs. Networks built with this architecture have a stronger ability to provide an efficient high-level feature representation. This is important for establishing an integrated speaker verification system which can be aware to potential spoofing attacks. As introduced in Section~\ref{sec_liter}, there are several shared ResBlocks for extracting discriminative representations, which are then used for both tasks. If we ignore the information and the content carried by a voice, a spoofed speech can be thought of as a Bona fide speech from a `fake' speaker. With this hypothesis, a spoofing detection task is similar to a two-speaker classification and a speaker verification task is a multi-speaker classification by its definition. Shared ResBlocks in the proposed multi-task learning network are used for capturing auxiliary relationships from both tasks. The experimental results in Section~\ref{sec_ERA} demonstrate the effectiveness and the benefit of this architecture. 

Another effective method employed to enhance the generalization of the proposed SR-ASV system is the MFM activation. For a multi-task learning based system, if the artifacts in spoofed speech are not properly handled, a biased result will be yielded. The commonly-used ReLU activation leads to a zero output value if a node is not active in the network. This might cause a loss of some information especially for the first several convolutional layers \cite{wu2018light}. By using the MFM activation, optimal feature at each location of different kernels are selected. A model with MFM can obtain a compact and robust representation while the gradients of MFM layers are sparse. Competitive nodes learning more generalized information can be output from the MFM layer. This property of MFM helps to force the network to extract effective embeddings for both tasks. 

By rebuilding and modifying the essential network layers like the MFM and residual blocks, the proposed network can be trained to extract speech embeddings of high discrimination and precise representation.  

\subsection{Multi-task learning}

The multi-task learning network is used in this paper to train the embedding extractor jointly. The tasks of spoofing detection and ASV are integrated at the model-level. The training space for the joint model can be expressed as: 
\begin{equation}
\label{eq_mtlspace}
{\rm H} = \left\{ {{\bm{X}_i},\bm{y}_i^{{\rm{SD}}},\bm{y}_i^{{\rm{ASV}}}} \right\}
\end{equation}
where $\bm{X}_i$ is the input of $i$th speaker. The $\bm{y}_i^{{\rm{SD}}}$ and $\bm{y}_i^{{\rm{ASV}}}$ are the corresponding one-hot encoded labels indicating the spoofing detection key and speaker ID, respectively. 

To optimize the joint model, we adopt the angular margin based softmax loss (A-softmax) which has been used for face recognition \cite{liu2017sphereface} and speaker embeddings extraction \cite{novoselov2018deep}. Recently, the A-softmax is also applied in spoofing detection to train deep learning based architectures \cite{lavrentyeva2019stc}. A-softmax is a well-regularized loss function by forcing learned features to be discriminative on a hypersphere manifold. This loss function can be described as:
\begin{strip}
	\begin{equation}
	\label{eq_asoftmax}
	{L_A}\left( {\bm{x},\bm{y},\bm{\theta} } \right) = \frac{1}{N}\sum\nolimits_i { - \log \left( {\frac{{{e^{\left\| {{x_i}} \right\|\cos \left( {m{\theta _{i,{y_i}}}} \right)}}}}{{{e^{\left\| {{x_i}} \right\|\cos \left( {m{\theta _{i,{y_i}}}} \right)}} + \sum\nolimits_{i \ne {y_i}} {{e^{\left\| {{x_i}} \right\|\cos \left( {m{\theta _{i,{y_i}}}} \right)}}} }}} \right)} 
	\end{equation}
\end{strip}
where $N$ is the number of training samples $\left\{ {{x_i}} \right\}_{i = 1}^N$ and labels $\left\{ {{y_i}} \right\}_{i = 1}^N$. The angle between a sample $x_i$ and the corresponding column $y_i$ of the fully connected classification layer weights $\bm{W}$ is denoted as ${{\theta _{i,{y_i}}}}$. In addition, $m$ is an integer that controls the size of an angular margin between classes. 

If we use ${L_{SD}}\left( {\bm{X},\bm{y}^{{\rm{SD}}}, \bm{\theta}^{{\rm{SD}}}} \right)$ and ${L_{ASV}}\left( {\bm{X},\bm{y}^{{\rm{ASV}}}, \bm{\theta}^{{\rm{ASV}}}} \right)$ to denote the loss functions for spoofing detection and ASV, we can obtain the total cost function as below for simplicity:
\begin{strip}
	\begin{equation}
	\label{eq_totalloss}
	J\left( {\bm{X},{\bm{y}^{{\rm{SD}}}},{\bm{y}^{{\rm{ASV}}}},{\rm{ }}{\bm{\theta} ^{{\rm{SD}}}},{\rm{ }}{\bm{\theta} ^{{\rm{ASV}}}}} \right) = {L_{SD}}\left( {\bm{X},\bm{y}^{{\rm{SD}}}, \bm{\theta}^{{\rm{SD}}}} \right) + {L_{ASV}}\left( {\bm{X},\bm{y}^{{\rm{ASV}}}, \bm{\theta}^{{\rm{ASV}}}} \right) + \frac{\lambda }{2}{\left\| \bm{W} \right\|^2}
	\end{equation}
\end{strip}
where the $\lambda$ is the regularization parameter which is optimized on the development subset.

\section{Experimental settings}
\label{sec_expsetting}

\subsection{Database}

To compare the proposed SR-ASV system with the state-of-the-art, it is evaluated on the ASVspoof 2017 and 2019 databases which are released by the challenge organizers\footnote{https://www.asvspoof.org/database}. Detailed statistical summaries of these two corpora can be found as in Table~\ref{tab:table_2017database} and Table~\ref{tab:table_2019database}.

\subsubsection{ASVspoof 2017 corpus}

The ASVspoof 2017 corpus has been used in the second challenge and was collected for countermeasures to replay spoofing attacks. Bona fide utterances are a sub-set of the RedDots corpus \cite{lee2015reddots} and spoofed utterances are the result of replaying and recording bona fide utterances using a variety of heterogeneous devices and acoustic environments. There are two versions for this database. Version 1.0 was used as the official corpus in the 2017 challenge while Version 2.0 was released afterwards. In this updated version, a number of data anomalies in the Version 1.0 of the corpus were removed to avoid potential influence on the detection results. In this paper we will utilize the Version 2.0 corpus to evaluate the proposed SR-ASV system in an impartial and objective manner. Furthermore, the protocol\footnote{https://www.asvspoof.org/index2017.html} used for the enrollment with regard to the ASVspoof 2017 corpus is the same as in \cite{Todisco2018}.

\subsubsection{ASVspoof 2019 corpus}

The other database is the ASVspoof 2019 corpus and it is designed for the third challenge. This database encompasses two partitions: logical access (LA) and physical access (PA) scenarios, which are all derived from the VCTK base corpus\footnote{http://dx.doi.org/10.7488/ds/1994}. All the spoofed speech are generated from the Bona fide data using diverse spoofing algorithms. More details of the spoofing approaches have been released in the challenge evaluation plan\footnote{https://www.asvspoof.org/asvspoof2019/asvspoof2019\_evaluation\_plan.pdf}.

There is another dataset included with the ASVspoof 2019 corpus and used for the enrollment of the baseline ASV system in the ASVspoof 2019 challenge. We also adopt this dataset in this paper to enroll the SR-ASV system. The details of the enrollment partition subset are given in Table~\ref{tab:table_2019databaseenroll}. All these speakers in the table are presented in the corresponding subsets in the ASVspoof 2019 corpus given in Table~\ref{tab:table_2019database}. Detailed numbers of speech samples from different genders are also provided for both partitions. 

\begin{table}[htbp]
	\renewcommand\arraystretch{1.3}
	\caption{Statistics of the ASVspoof 2017 Version 2.0 database.}
	\label{tab:table_2017database}
	\centering
	\begin{tabular}{lccccc}
		\toprule
		\multirow{2}{*}{Subset}&
		\multirow{2}{*}{\#Speaker}&
		\multirow{2}{*}{\tabincell{c}{\#Replay\\sessions}}&
		\multirow{2}{*}{\tabincell{c}{\#Replay\\Config.}}& 
		\multicolumn{2}{c}{\#Utterances}\\  
		\cmidrule(lr){5-6}  
		&&&&Bona fide&Replay\\
		\midrule
		Training     & 10  & 6   & 3  & 1507 & 1507         \\
		Dev.         & 8   & 10  & 10 & 760  & 950          \\
		Eval.        & 24  & 161 & 57 & 1298 & 12008        \\
		\hline
		Total        & 42  & 177 & 61 & 3565 & 14465        \\
		\bottomrule
	\end{tabular}
\end{table}
\begin{table}[htbp]
	\renewcommand\arraystretch{1.2}
	\caption{Statistics of the ASVspoof 2019 database.}
	\label{tab:table_2019database}
	\centering
	\begin{tabular}{l|c|c|cc|cc}
		\toprule
		\multirow{3}{*}{Subset}&\multicolumn{2}{c}{\#Speaker}&\multicolumn{4}{c}{\#Utterances}\\  
		\cmidrule(lr){2-3} \cmidrule(lr){4-7}  
		&\multirow{2}{*}{M}&\multirow{2}{*}{F}&\multicolumn{2}{c}{LA}&\multicolumn{2}{c}{PA}\\
		\cmidrule(lr){4-5} \cmidrule(lr){6-7}
		&&&Bona fide&Spoof&Bona fide&Spoof\\
		\midrule
		Training        & 8  & 12 & 2580 & 22800 & 5400  & 48600  \\
		Dev.            & 8  & 12 & 2548 & 22296 & 5400  & 24300  \\
		Eval.           & 21 & 27 & 7355 & 63882 & 18090 & 116640 \\
		\hline
		Total                  & 37 & 51 & 12483& 108978& 28890 & 189540 \\
		\bottomrule
	\end{tabular}
\end{table}
\begin{table}[htbp]
	\renewcommand\arraystretch{1.3}
	\caption{Statistics of the enrollment partition of the ASVspoof 2019 database.}
	\label{tab:table_2019databaseenroll}
	\centering
	\begin{tabular}{l|c|c|cc|cc}
		\toprule
		\multirow{3}{*}{Subset}&\multicolumn{2}{c}{\#Speaker}&\multicolumn{4}{c}{\#Utterances}\\  
		\cmidrule(lr){2-3} \cmidrule(lr){4-7}  
		&\multirow{2}{*}{M}&\multirow{2}{*}{F}&\multicolumn{2}{c}{LA}&\multicolumn{2}{c}{PA}\\
		\cmidrule(lr){4-5} \cmidrule(lr){6-7}
		&&&M&F&M&F\\
		\midrule
		Enrollment\_Dev.           & 4  & 6  & 76  & 66  & 2052   & 1782  \\
		Enrollment\_Eval.          & 21 & 27 & 399 & 297 & 10773  & 8019  \\
		\hline
		Total                      & 25 & 33 & 475 & 363 & 12825  & 9801  \\
		\bottomrule
	\end{tabular}
\end{table}
Similar to the baseline ASV system \cite{todisco2019asvspoof} used in the challenge, the VoxCeleb corpus\footnote{An audio-visual dataset contains over 100K utterances for 1251 celebrities, which are extracted from videos uploaded to YouTube. This dataset is available at http://www.robots.ox.ac.uk/$\sim$vgg/data/voxceleb/. } is used to pre-train the shared hidden layers of the MTL architecture used in this work. Then the entire network is trained with the training subsets of the challenge corpora. PLDA adaptation is performed with the enrollment subsets, which include disjoint, Bona fide, and in-domain speech samples.

\subsection{Baseline and benchmark systems}
\label{sec_BSsystems} 

For spoofing detection, we adopt several cepstral coefficients based systems, including the Constant Q Cepstral Coefficient (CQCC), the Linear Frequency Cepstral Coefficient (LFCC), the Short-time Fourier Transform Cepstral Coefficient (SFTCC) and the Inverted Mel-Frequency Cepstral Coefficient (IMFCC). Another kind of acoustic features used in benchmark systems is time-frequency representation based features. A summary list of all the spoofing detection systems is given in Table~\ref{tab:table_baselinesummary}. The BS1 and BS2 are two baseline systems while the BS3-BS6 are four top performing benchmark systems submitted to the ASVspoof 2019 challenge.
\begin{table*}[htbp]
	\renewcommand\arraystretch{1.2}
	\caption{Summary of the baseline spoofing detection systems with implementation details.}
	\label{tab:table_baselinesummary}
	\centering
	\begin{tabular}{|l|l|l|m{8cm}|c|}
		\toprule
		\textbf{Systems}&\textbf{Front-ends}&\textbf{Back-ends}&\textbf{Implementation Details}&\textbf{Refs.}\\
		\midrule
		BS1 & CQCC & GMM  & $s+\Delta+\Delta\Delta = 90$/No. of bins = 96/Re-sampling period = 16  & \cite{todisco2017constant} \\
		\hline
		BS2 & LFCC & GMM  & $s+\Delta+\Delta\Delta = 60$/No. of filters = 20/frame size = 20ms/frame shift = 10ms/FFT bins = 512  & \cite{alegre:hal-00849138} \\
		\hline
		BS3 & STFCC& GMM  & $s+\Delta+\Delta\Delta = 60$/frame size = 20ms/frame shift = 10ms/FFT bins = 512  & - \\
		\hline
		BS4 & IMFCC& GMM  & $s+\Delta+\Delta\Delta = 60$/No. of filters = 20/frame size = 20ms/frame shift = 10ms/FFT bins = 512  & \cite{sahidullah2015comparison} \\
		\hline
		BS5 & FFT  & LCNN & unified size of $864 \times 400$/window length = 1724/step = 0.0081/Blackman window  & \cite{lavrentyeva2017audio,lavrentyeva2019stc} \\
		\hline
		BS6 & CQT  & LCNN & unified size of $864 \times 400$/No. of bins = 96/Re-sampling period = 16  & \cite{lavrentyeva2017audio,lavrentyeva2019stc} \\
		\bottomrule
	\end{tabular}
\end{table*}
\begin{figure}[htbp]
	\centering
	\includegraphics[width=8cm,height=10cm]{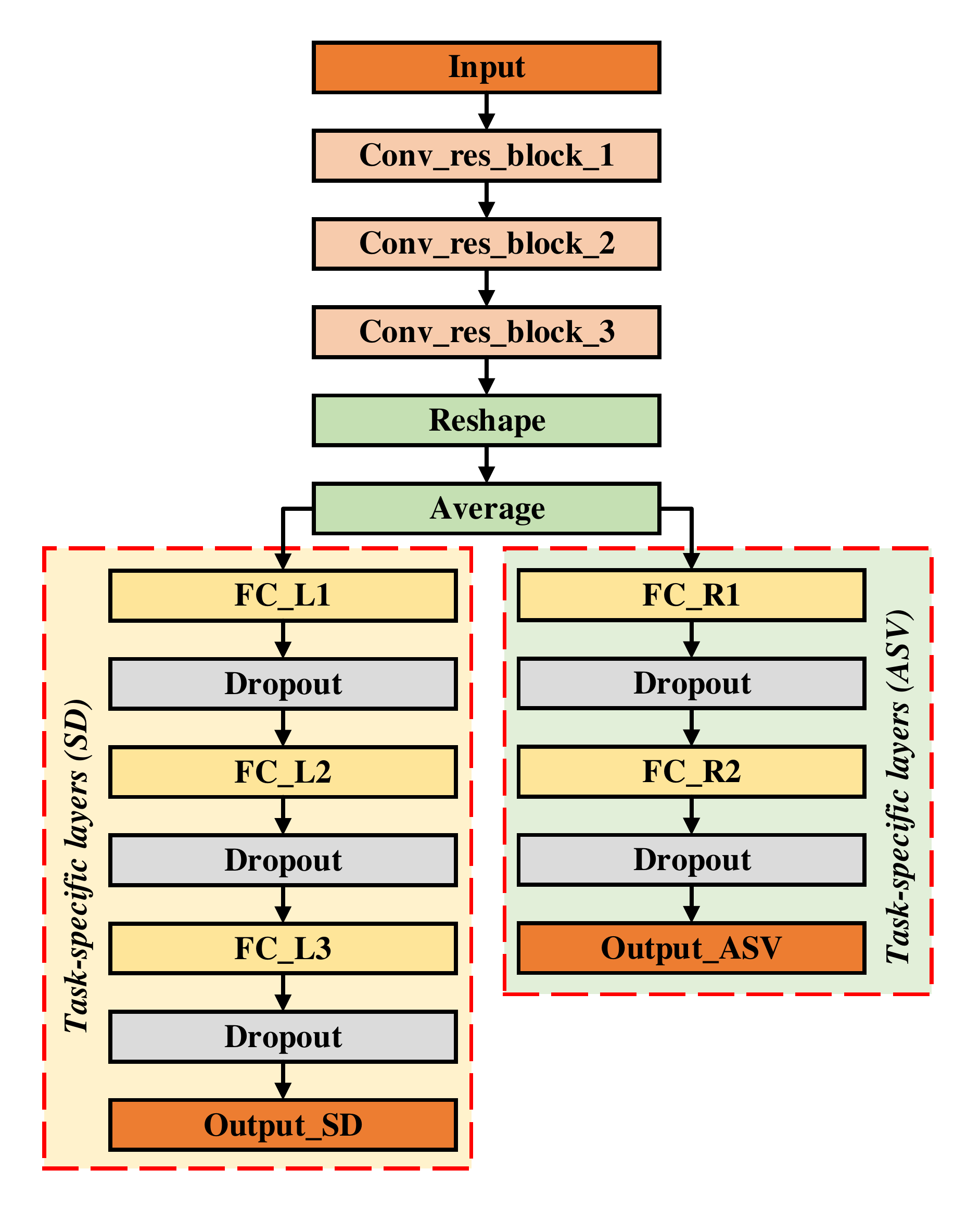}
	\caption{The multi-task learning architecture used in the SR-ASV system.}
	\label{fig:modelSRASV}
\end{figure}
\begin{figure}[htbp]
	\centering
	\includegraphics[width=8cm,height=5.5cm]{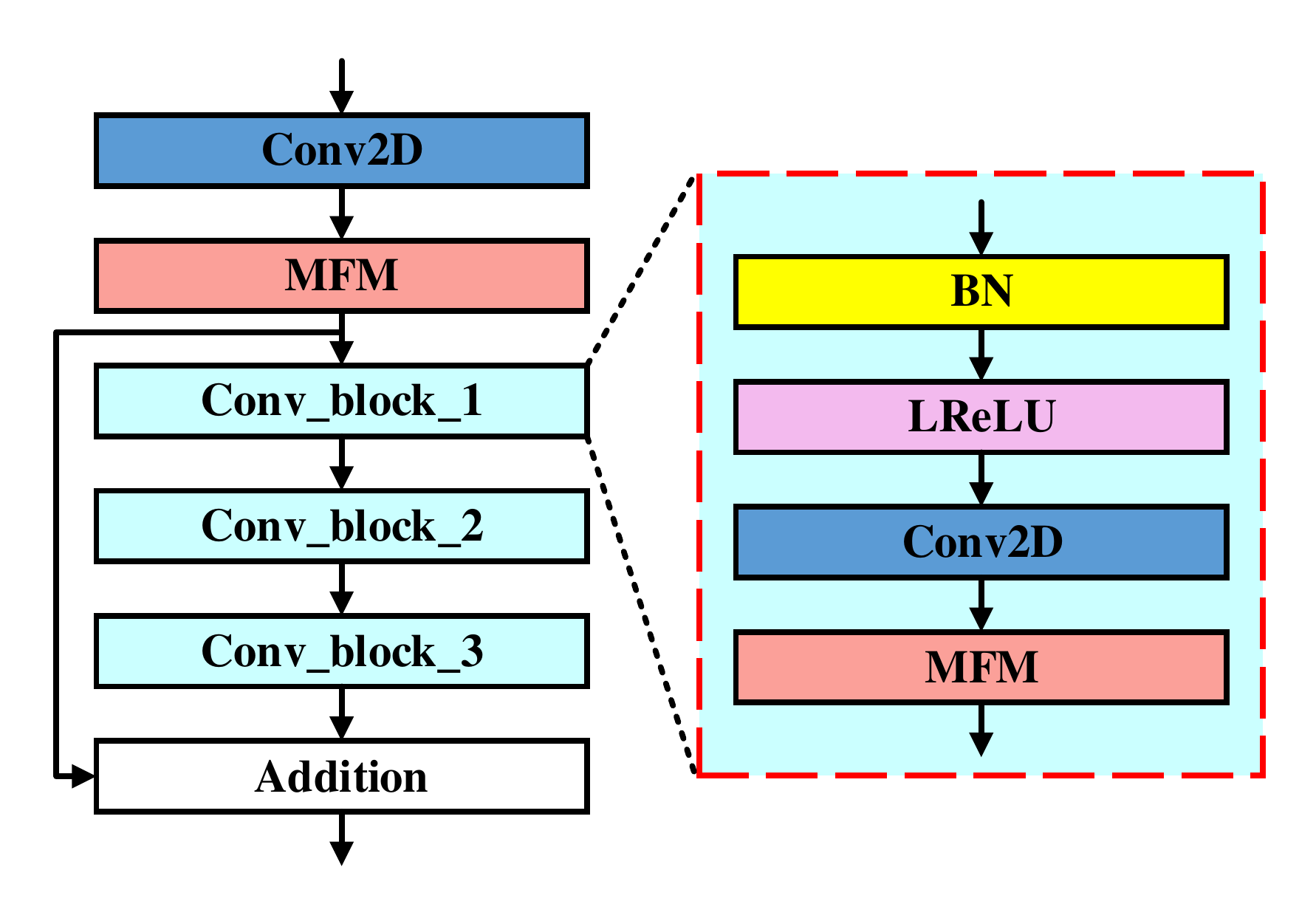}
	\caption{The residual convolutional block used in the MTL network.}
	\label{fig:modelSRASV_crb}
\end{figure}

The baseline ASV system used to compare is from the ASVspoof 2019 challenge \cite{todisco2019asvspoof}. It uses x-vector speaker embeddings \cite{snyder2018x} together with a PLDA back-end \cite{prince2007probabilistic}. More detailed configurations for the baseline ASV system are given in \cite{lavrentyeva2019stc,todisco2019asvspoof}.

\subsection{Front-end processing}
\label{sec_prepro}

Two types of time-frequency representation based speech features including the constant Q transformation (CQT) and log linear-filterbank (LLFB) are adopted in this work. To extract the CQT-TFR feature for each utterance, we apply the open-source Matlab toolkit\footnote{http://audio.eurecom.fr/content/software}. The maximum and the minimum frequency in the constant Q transform are set as ${F_{\max }} = {F_{sample}}/2$ and ${F_{\min }} = {F_{\max }}/{2^9}$ respectively. The Nyquist frequency of the database is ${F_{sample}} = 16{\rm{kHz}}$. The number of octaves is 9 and the number of bins per octave B is set to 96, which results in a time shift of 8 ms. The parameter $\gamma $ is set to $\gamma  = \Gamma  = 228.7 * \left( {{2^{\left( {1/B} \right)}} - {2^{\left( { - 1/B} \right)}}} \right)$. The re-sampling period is $d = 16$. After the CQT is applied on an utterance, we take the same truncating process as in \cite{lavrentyeva2017audio,lavrentyeva2019stc} to get spectral features with a size of $864 \times 400$. LLFB features are extracted by a classical pipeline for filterbank based features. By truncating the utterance-level features with the same processing used for the CQT feature, LLFB features with unified shapes of $80 \times 400$ (where 400 is the number of time frames) can be obtained. 

\subsection{Network architectures and configurations}

The shared hidden layers in the MTL network are built by a series of residual convolutional blocks. Separate task-specific fully connected layers are used after the shared hidden layers. The architecture of the entire network\footnote{The corresponding codes used for reimplementing the experiments will be released later.} is demonstrated in Fig.\ref{fig:modelSRASV}. Due to the relatively large dimension of speech features and the limited computing power, in each block we apply the Max-Feature-Map (MFM) activations to simplify the deep learning architecture. The MFM activation adopts a competitive relationship to obtain a compact representation and performs feature filter selection. More details can be found in \cite{wu2018light}.

\begin{table}[htbp]
	\renewcommand\arraystretch{1.2}
	\caption{An example of the network configurations of the MTL network. (Statistics of the task-specific layers for ASV are given in brackets.)}
	\label{tab:table_SRASV}
	\centering
	\begin{tabular}{lllr}
		\toprule
		\textbf{Type}&\textbf{Output}&\textbf{\#Params}\\
		\midrule
		Input                &$864 \times 400 \times 1 $ &-           \\
		\hline
		Conv\_res\_block\_1  &$432 \times 200 \times 16$ &6K          \\
		Conv\_res\_block\_2  &$216 \times 100 \times 32$ &32K         \\
		Conv\_res\_block\_3  &$108 \times 50  \times 64$ &128K        \\
		\hline
		Reshape              &$108 \times 3200$          &-           \\
		Average              &3200                       &-           \\
		\hline
		FC                   &512 (512)                  &1.6M (1.6M) \\
		Dropout              &512 (512)                  &-           \\
		FC                   &128 (128)                  &65K  (65K)  \\
		Dropout              &128 (128)                  &-           \\
		FC                   &64  (-)                    &8K   (-)    \\
		Dropout              &64  (-)                    &-           \\
		\hline
		Output               &2   (78)                   &130  (10K)  \\
		\hline
		Total                &-                          &3.6M   \\
		\bottomrule
	\end{tabular}
\end{table}

\begin{table}[htbp]
	\renewcommand\arraystretch{1.3}
	\caption{Configurations of residual convolutional blocks used in the MTL network.}
	\label{tab:table_crblock_MTL}
	\centering
	\begin{tabular}{lllr}
		\toprule
		\textbf{Type}&\textbf{Filter / Stride}&\textbf{Output}&\textbf{\#Params}\\
		\midrule
		Conv2D               &$3 \times 3$ / $2 \times 2$ &$432 \times 200 \times 32$ &320   \\
		MFM                  &-                           &$432 \times 200 \times 16$ &-     \\
		\hline
		BatchNormalization   &-                           &$432 \times 200 \times 16$ &64    \\
		LReLU                &-                           &$432 \times 200 \times 16$ &-     \\
		Conv2D               &$1 \times 1$ / $1 \times 1$ &$432 \times 200 \times 32$ &544   \\
		MFM                  &-                           &$432 \times 200 \times 16$ &-     \\
		\hline
		BatchNormalization   &-                           &$432 \times 200 \times 16$ &64    \\
		LReLU                &-                           &$432 \times 200 \times 16$ &-     \\
		Conv2D               &$3 \times 3$ / $1 \times 1$ &$432 \times 200 \times 32$ &4640  \\
		MFM                  &-                           &$432 \times 200 \times 16$ &-     \\
		\hline
		BatchNormalization   &-                           &$432 \times 200 \times 16$ &64    \\
		LReLU                &-                           &$432 \times 200 \times 16$ &-     \\
		Conv2D               &$1 \times 1$ / $1 \times 1$ &$432 \times 200 \times 32$ &544   \\
		MFM                  &-                           &$432 \times 200 \times 16$ &-     \\
		\hline
		Addition             &-                           &$432 \times 200 \times 16$ &-     \\
		\hline
		Total                &-                           &-                          &6K    \\
		\bottomrule
	\end{tabular}
\end{table}
\begin{table}[htbp]
	\renewcommand\arraystretch{1.2}
	\caption{t-DCF cost function parameters assumed in ASVspoof 2019.}
	\label{tab:table_tdcfparameters}
	\centering
	\begin{tabular}{ccccccc}
		\toprule
		\multicolumn{3}{c}{\textbf{Priors}}&
		\multicolumn{2}{c}{\textbf{SD costs}}&
		\multicolumn{2}{c}{\textbf{ASV costs}}\\  
		\cmidrule(lr){1-3} \cmidrule(lr){4-5} \cmidrule(lr){6-7} 
		${\pi _{{\rm{tar}}}}$&${\pi _{{\rm{non}}}}$&${\pi _{{\rm{spoof}}}}$&$C_{{\rm{miss}}}^{{\rm{SD}}}$&$C_{{\rm{fa}}}^{{\rm{SD}}}$&$C_{{\rm{miss}}}^{{\rm{ASV}}}$&$C_{{\rm{fa}}}^{{\rm{ASV}}}$\\
		\midrule
		0.9405&0.0095&0.05&1&10&1&10 \\
		\bottomrule
	\end{tabular}
\end{table}
As shown in Fig.\ref{fig:modelSRASV}, there are three residual convolutional blocks used in the MTL network, of which the outputs are reshaped and averaged before the task-specific layers. The detailed components of the residual convolutional block are shown in Fig.\ref{fig:modelSRASV_crb}. For the whole network, we use 6 convolutional layers, 6 Network in Network (NIN) layers, 12 MFM layers and 5 fully connected layers. We modify the connections in the residual blocks to perform the full pre-activation for the optimal gradient flow \cite{he2016deep}. For the shared residual convolutional blocks, we insert the Leaky ReLU (LReLU) activation function and batch normalization to stabilize the model training. The number of filters used in the three residual convolutional blocks are 32, 64 and 128, respectively.

Detailed configurations and statistics of the network parameters are listed in Table~\ref{tab:table_SRASV} and Table~\ref{tab:table_crblock_MTL}. The He normal initializer \cite{he2015delving} and the L2 regularizer are used with a regularization parameter $\lambda = 0.001$. To avoid the over-fitting issue, dropout 0.7 is used in the network. The Adam optimizer is adopted. Class weights are needed to address the imbalanced training data.

\subsection{Performance metrics}
\subsubsection{Tandem decision cost function (t-DCF)}

In \cite{Kinnunen2018}, the tandem decision cost function (t-DCF) was proposed for the assessment of combined spoofing countermeasures and ASV. The t-DCF has been adopted as the official primary performance metric\footnote{www.asvspoof.org/asvspoof2019/asvspoof2019\_evaluation\_plan.pdf}. The basic form of t-DCF is expressed as below:

\begin{equation}
\label{eq_tdcf}
{\rm{t - DCF}}\left( s \right) = {C_1}P_{{\rm{miss}}}^{{\rm{SD}}}\left( s \right) + {C_2}P_{{\rm{fa}}}^{{\rm{SD}}}\left( s \right) + {C_0}
\end{equation}
where $P_{{\rm{miss}}}^{{\rm{SD}}}\left( s \right)$ and $P_{{\rm{fa}}}^{{\rm{SD}}}\left( s \right)$ are the false rejection rate (FRR) and the false alarm rate (FAR) of the spoofing detection system at threshold $s$, respectively. The constants $C_0$, $C_1$ and $C_2$ can be calculated with the t-DCF costs, priors and the ASV system detection errors:
\begin{equation}
\label{eq_c0c1c2}
\left\{ \begin{array}{l}
{C_0} = {\pi _{{\rm{tar}}}}C_{{\rm{miss}}}^{{\rm{ASV}}}P_{{\rm{miss}}}^{{\rm{ASV}}} + {\pi _{{\rm{non}}}}C_{{\rm{fa}}}^{{\rm{ASV}}}P_{{\rm{fa}}}^{{\rm{ASV}}}\\
{C_1} = {\pi _{{\rm{tar}}}}C_{{\rm{miss}}}^{{\rm{SD}}} - {C_0}\\
{C_2} = C_{{\rm{fa}}}^{{\rm{SD}}}{\pi _{{\rm{spoof}}}}\left( {1 - P_{{\rm{miss,spoof}}}^{{\rm{ASV}}}} \right)
\end{array} \right.
\end{equation}

In (\ref{eq_c0c1c2}), $C_{{\rm{miss}}}^{{\rm{SD}}}$, $C_{{\rm{fa}}}^{{\rm{SD}}}$, $C_{{\rm{miss}}}^{{\rm{ASV}}}$ and $C_{{\rm{fa}}}^{{\rm{ASV}}}$ are the costs of the spoofing detection and ASV systems respectively for rejection (miss) of a positive (Bona fide or target) trial and false acceptance (fa) of a negative (spoof or nontarget) trial. Furthermore, we assert a priori probabilities of target (${\pi _{{\rm{tar}}}}$), nontarget (${\pi _{{\rm{non}}}}$) and spoof (${\pi _{{\rm{spoof}}}}$) classes. Note that ${\pi _{{\rm{tar}}}} + {\pi _{{\rm{non}}}} + {\pi _{{\rm{spoof}}}} = 1$. In this work, we adopt the same costs and prior probabilities used in the ASVspoof 2019 challenge, which are shown in Table~\ref{tab:table_tdcfparameters}.

In this work, the normalized t-DCF is adopted as the primary performance metric and given as below: 
\begin{equation}
\label{eq_tdcfnorm}
{\rm{t - DC}}{{\rm{F}}_{{\rm{norm}}}}\left( s \right) = \frac{{{C_1}}}{{{C_2}}}P_{{\rm{miss}}}^{{\rm{SD}}}\left( s \right) + P_{{\rm{fa}}}^{{\rm{SD}}}\left( s \right)
\end{equation}
\begin{table}[htbp]
	\renewcommand\arraystretch{1.2}
	\caption{EERs(\%) and t-DCF comparison between different integrated systems on the evaluation subset of the ASVspoof 2017 Version 2.0 corpus. ASV denotes the speaker verification and SD denotes the spoofing detection task. Results not provided by the authors are denoted by `-'.  }
	\label{tab:table_2017results}
	\centering
	\begin{tabular}{lcccc}
		\toprule
		\multirow{2}{*}{Systems}&
		\multirow{2}{*}{ASV}&
		\multicolumn{2}{c}{SD}&
		\multirow{2}{*}{Average}\\  
		\cmidrule(lr){3-4} 
		&&EER&t-DCF&\\
		\midrule
		CQCC-GMM \cite{Todisco2018}       & 4.71 & 18.11 &-& 11.41 \\
		Joint decision \cite{li2020joint} & -    & -     &-& 8.97  \\
		\hline
		SR-ASV (CQT)     & 3.16 & 8.22 &0.2022& 5.69  \\
		SR-ASV (LLFB)    & 3.02 & 8.76 &0.2173& 5.89  \\
		\hline
		Fusion           & \textbf{2.87} & \textbf{8.05}  &\textbf{0.1974}& \textbf{5.46}   \\
		\bottomrule
	\end{tabular}
\end{table}
Note that in the ASVspoof 2019 challenge, the decision scores for the baseline ASV system were not made available during the evaluation. The final t-DCFs of every spoofing detection system were calculated by the organizers with the spoofing detection scores submitted from all teams and the ASV scores pre-calculated. In this work, the t-DCFs of the proposed SR-ASV system are computed by the resultant ASV and spoofing detection scores and used for assessing the performance of spoofing detection.

\subsubsection{Equal error rate (EER)}

The EER is used to predetermine the threshold values for the FAR and the FRR. When these two rates are equal, the corresponding value is referred to as the EER. The lower the EER value, the higher the accuracy of a biometric system. In this work, EER is used to measure the system performance for both the spoofing detection and the ASV tasks. 

\section{Experimental results and analysis}
\label{sec_ERA}

\subsubsection{Results on the ASVspoof 2017 corpus}

The comparison between different integrated systems on the evaluation subset of the ASVspoof 2017 corpus is given in Table~\ref{tab:table_2017results}. The benchmark systems adopt both the conventional method like the Gaussian back-end fusion \cite{Todisco2018} and the multi-task learning approach based on contrastive loss \cite{li2020joint}. For the ASV task, our proposed SR-ASV system with the LLFB feature achieves the best EER of 3.02\%. Furthermore, for the spoofing detection system, the proposed system with the CQT feature can provide a low EER of 8.22\% and a t-DCF of 0.2022 on the evaluation subset. The results can be further improved by fusing the proposed systems with the Bosaris toolkit\footnote{https://sites.google.com/site/bosaristoolkit/home}. From the fusion of the SR-ASV systems using CQT and LLFB features, an averaged EER of 5.46\% can be obtained.

\subsubsection{Results on the ASVspoof 2019 corpus}

\begin{table}[htbp]
	\renewcommand\arraystretch{1.2}
	\caption{The \MakeTextLowercase{t}-DCF and EER results for baseline systems and the proposed SR-ASV system on the LA partition for the spoofing detection task.}
	\label{tab:table_LA_results}
	\centering
	\begin{tabular}{l|c|c|c|c}
		\toprule
		\multirow{2}{*}{Systems}&
		\multicolumn{2}{c}{Dev.}&
		\multicolumn{2}{c}{Eval.}\\  
		\cmidrule(lr){2-3} \cmidrule(lr){4-5} 
		&t-DCF&EER&t-DCF&EER\\
		\midrule
		AS1 (CQT)      & 0.0003 & 0.03 & 0.1127 & 5.04 \\
		AS2 (LLFB)     & 0.0057 & 0.19 & 0.1863 & 6.92 \\
		\hline
		BS1 (CQCC-GMM) & 0.0123 & 0.43 & 0.2366 & 9.57 \\
		BS2 (LFCC-GMM) & 0.0663 & 2.71 & 0.2116 & 8.09 \\
		BS3 (STFCC-GMM)& 0.0000 & 0.00 & 0.1400 & 5.97 \\
		BS4 (IMFCC-GMM)& 0.0002 & 0.03 & 0.2198 & 9.49 \\
		BS5 (FFT-LCNN) & 0.0009 & 0.04 & 0.1028 & \textbf{4.53} \\
		BS6 (CQT-LCNN) & 0.0000 & 0.00 & 0.1014 & 4.58 \\
		\hline
		SR-ASV (CQT)   & 0.0000 & 0.00 & \textbf{0.1009} & 4.55 \\
		SR-ASV (LLFB)  & 0.0005 & 0.02 & 0.1034 & 4.60 \\
		\hline
		Fusion         & 0.0000 & 0.00 & \textbf{0.0517} & \textbf{1.73}  \\
		\bottomrule
	\end{tabular}
\end{table}
\begin{table}[htbp]
	\renewcommand\arraystretch{1.2}
	\caption{The \MakeTextLowercase{t}-DCF and EER results for baseline systems and the proposed SR-ASV system on the PA partition for the spoofing detection task. Results not provided by the authors are denoted by `-'.}
	\label{tab:table_PA_results}
	\centering
	\begin{tabular}{l|c|c|c|c}
		\toprule
		\multirow{2}{*}{Systems}&
		\multicolumn{2}{c}{Dev.}&
		\multicolumn{2}{c}{Eval.}\\  
		\cmidrule(lr){2-3} \cmidrule(lr){4-5} 
		&t-DCF&EER&t-DCF&EER\\
		\midrule
		AS1 (CQT)      & 0.0522 & 1.50 & 0.7521 & 3.29 \\
		AS2 (LLFB)     & 0.0661 & 3.31 & 0.1380 & 5.35 \\
		\hline
		BS1 (CQCC-GMM) & 0.1953 & 9.87  & 0.2454 & 11.04 \\
		BS2 (LFCC-GMM) & 0.2554 & 11.96 & 0.3017 & 13.54 \\
		BS3 (STFCC-GMM)& 0.1462 & 7.09  & 0.2129 & 9.07  \\
		BS4 (IMFCC-GMM)& 0.1464 & 7.04  & 0.2128 & 9.04  \\
		Joint decision \cite{li2020joint} &- &- &- & 8.55 \\
		BS5 (FFT-LCNN) & 0.0759 & 3.92  & 0.6713 & 2.75  \\
		BS6 (CQT-LCNN) & 0.0197 & 0.80  & 0.0295 & 1.23  \\
		\hline
		SR-ASV (CQT)   & 0.0186 & 0.73  & \textbf{0.0287} & \textbf{1.15}  \\
		SR-ASV (LLFB)  & 0.0346 & 2.02  & 0.3597 & 1.62  \\
		\hline
		Fusion         & 0.0005 & 0.03  & \textbf{0.0108} & \textbf{0.55}  \\
		\bottomrule
	\end{tabular}
\end{table}
Since the ASVspoof 2019 corpus is newly released and contains both the logical and physical access attacks, we investigate more on this general database. The performance of our proposed SR-ASV and other state-of-the-art systems for the spoofing detection task are presented in Table~\ref{tab:table_LA_results} and Table~\ref{tab:table_PA_results} for the LA and PA partitions, respectively. The BS1 and BS2 are two baseline systems used in the 2019 challenge. In addition to that, the BS3 to BS6 are four top performing benchmark systems submitted to the challenge. Detailed introductions of these systems have been given in section~\ref{sec_BSsystems}.

\begin{figure*}[htbp]
	\centering
	\includegraphics[width=15cm,height=8cm]{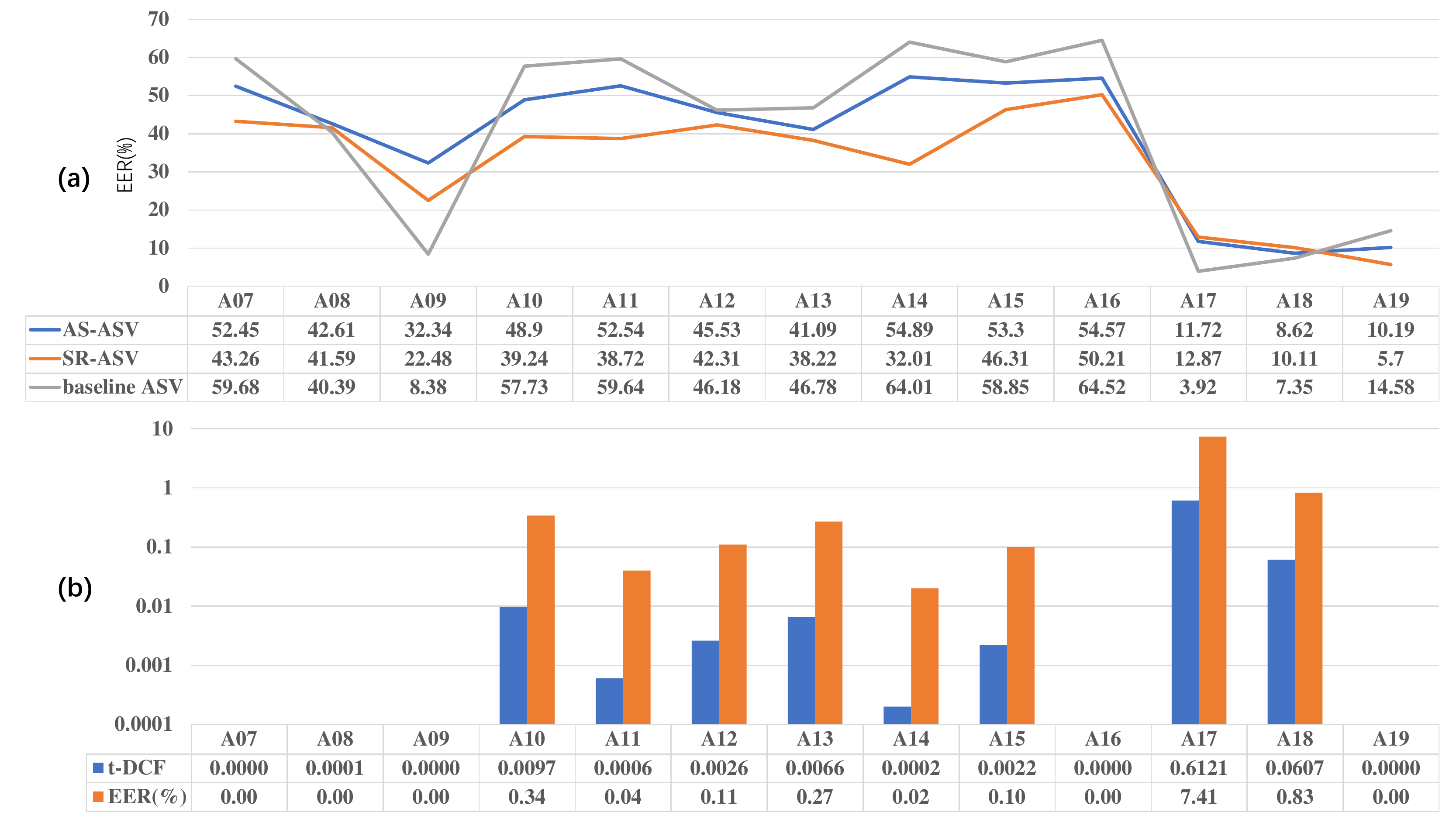}
	\caption{Performance of ASV systems across varying LA spoofing attack types in the evaluation subset. In (a), the curves indicate the resultant EERs when the SR-ASV and the challenge baseline ASV system are under attack and no anti-spoofing detectors are used. In (b), the t-DCFs and EERs of the SR-ASV system on different LA attacks are provided.  }
	\label{fig:LAattacks}
\end{figure*}
\begin{figure*}[htbp]
	\centering
	\includegraphics[width=15cm,height=8cm]{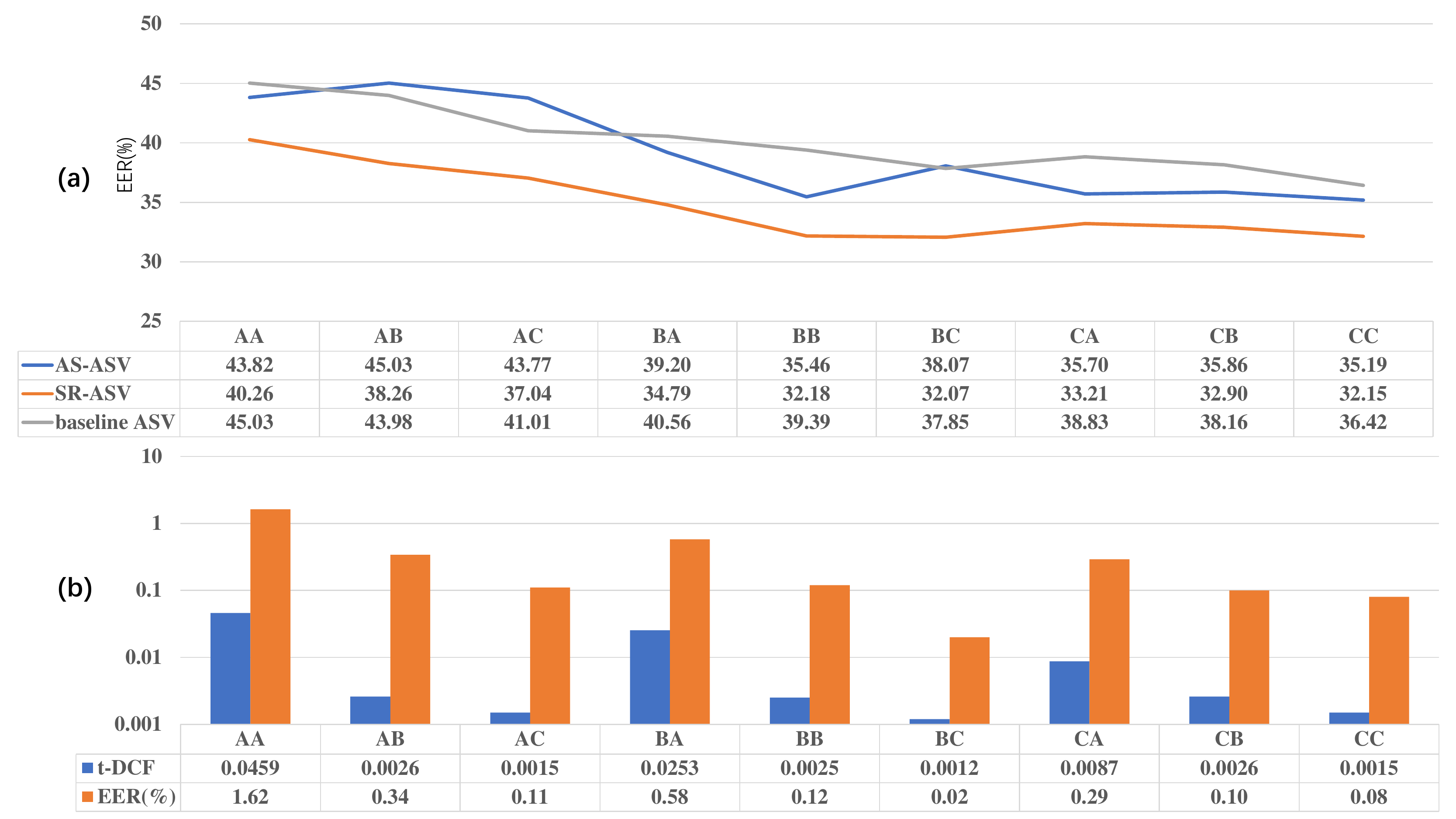}
	\caption{Performance of ASV systems across varying PA spoofing attack types in the evaluation subset. In (a), the curves indicate the resultant EERs when the SR-ASV and the challenge baseline ASV system are under attack and no anti-spoofing detectors are used. In (b), the t-DCFs and EERs of the SR-ASV system on different PA attacks are provided.  }
	\label{fig:PAattacks}
\end{figure*}
Compared to the traditional cepstral coefficients, experiment results for deep learning based systems display a better detection ability and verification accuracy. This proves that a more comprehensive and discriminative representation can be learned by deep learning based architectures with fine-tuned parameters. The best t-DCF results of single system on both partitions are obtained by the SR-ASV system using CQT-TFR features. Moreover, the performance can be further improved by fusing the SR-ASV (CQT) and SR-ASV (LLFB) scores. The corresponding fusion results are listed in the last rows in Table~\ref{tab:table_LA_results} and Table~\ref{tab:table_PA_results}. Specifically, for the LA partition, the t-DCFs of baseline systems (BS1 and BS2) used in the challenge are reduced almost 80\%. Furthermore, a significant improvement is also observed for the PA partition. The lowest t-DCF and EER on the evaluation subset of the PA partition are 0.0108 and 0.55, respectively.

Detailed performance comparison of the best single SR-ASV system and the challenge baseline ASV system across all spoofing attacks in the evaluation subsets are illustrated in Fig.\ref{fig:LAattacks} and Fig.\ref{fig:PAattacks} for the LA and PA partitions, respectively. Note that the results shown are all from single ASV systems. All the results of the challenge baseline ASV system are released in \cite{todisco2019asvspoof}.

To investigate if the proposed system efficiently alleviates the negative influence from spoofing attacks to ASV, we present the EER results when the SR-ASV and the challenge baseline ASV system are under attack while no external anti-spoofing systems are deployed. The curves of the verification performance are given in (a) of Fig.\ref{fig:LAattacks} and Fig.\ref{fig:PAattacks}. As shown in (a) of Fig.\ref{fig:LAattacks}, for the LA scenario there are several spoofing attacks that degrade ASV systems heavily and are difficult to detect, such as attacks A10, A13 and A15. They are all based on neural waveform or waveform concatenation skills. The proposed SR-ASV system performs worse than the baseline ASV system on the A09, A17 and A18 attacks, which are mounted with vocoder and waveform filtering. This is probably caused by the different embedding extraction approaches used in the networks. In the baseline ASV system, the x-vector based embedding is extracted from a statistical pooling layer with frame-level representations, while in our proposed system the embeddings are generated from utterance-level input features. The different perceptibilities of these embeddings lead to different sensibilities depending on the type of LA spoofing techniques. Interestingly, A17, a VAE-based voice conversion with waveform filtering, poses little threat to ASV systems while it is the most difficult to detect (as seen in (b) of Fig.\ref{fig:LAattacks}). This indicates that the spoofed speech generated from A17 is more of a threat as it tends to be verified as genuine target samples by the ASV systems under attack. We also make the observation that, the different detection performance across the different types of attacks motivates the demand for more research on generalized detection systems at the forefront of cutting-edge spoofing technology. 

Compared to the LA partition, the EER curves of these two ASV systems in (a) of Fig.\ref{fig:PAattacks} on the PA scenario are more consistent and stable. The performance gap between tasks of spoofing detection and speaker verification is fixed across all the nine replay configurations, i.e. replay attacks. When high-quality replay attacks are mounted, the EERs of these two ASV systems increase expectedly. This confirms that the quality of replay attack is the principal factor to be considered in anti-spoofing countermeasures. The overall performance of the SR-ASV system is better than the baseline ASV system. This indicates that the proposed integrated framework can improve the resistance to spoofing attacks for ASV. 

To assess the spoofing detection performance of the proposed integrated system on different attacks, the resultant t-DCFs and EERs are given in (b) of Fig.\ref{fig:LAattacks} and Fig.\ref{fig:PAattacks}. For cases of some unknown spoofing attacks in the LA partition, the performance is degraded at varying levels as shown in (b) of Fig.\ref{fig:LAattacks}. The proposed system performs poorly especially for the A17 and A18 spoofing attacks, which apply voice conversion with waveform filtering and vocoder. In contrast, for spoofed samples generated by speech synthesis techniques (A07-A12), relatively stable results are achieved. 

In (b) of Fig.\ref{fig:PAattacks}, a clear trend on the t-DCF and EER can be seen with different replay attacks. The system performance is mainly affected by the replay configurations which include replay device quality, distances to the original speakers and to ASV system and reverberation characteristics. Poor system performance is observed on high-quality replay attacks. This type of attack is mounted by using high-quality loudspeakers and recorders at a small distance to talkers in a quiet room. 

\subsubsection{Ablative study for the proposed system}

To explore the cause of the performance gains achieved and for a more comprehensive justification of the proposed SR-ASV system, we present the results for single spoofing detection systems in Table~\ref{tab:table_LA_results} and Table~\ref{tab:table_PA_results}. The systems in the first two rows with name of AS1 (CQT) and AS2 (LLFB) are two spoofing detection systems for ablative study. The backbone network used in either system is similar to the proposed SR-ASV system. However, the task-specific layers for the ASV task are pruned to remove the benefits of MTL on the speaker verification. By comparing the AS1 and AS2 with SR-ASV systems, it is clearly shown that the embeddings extracted when the task-specific layers for ASV are used can help to further improve the performance on anti-spoofing. There is a 10\%-30\% improvement on both the EER and t-DCF scores. 

We also give an ablative study on the ASV task. By using the same pruning manner on the backbone MTL, we cut the task-specific layers for the spoofing detection task to remove any influence on the ASV. The results of this single ablative study ASV system (AS-ASV) are shown in Fig.\ref{fig:LAattacks} (a) and Fig.\ref{fig:PAattacks} (a). Without the auxiliary information provided from the spoofing detection task, there is a clear performance gap between the AS-ASV and the SR-ASV systems. The AS-ASV system performs consistently worse compared to the proposed SR-ASV system on different types of spoofing attacks. Despite the steady performance on most other attacks, the increased EERs on the A17 and A18 still indicate the limitations of the proposed model for spoofed speech generated by novel voice conversion techniques.  

\section{Conclusion}
\label{sec_end}

In this paper, we explored to investigate on integrated spoofing-robust speaker verification systems which can effectively resist varying attacks. We proposed a multi-task learning based framework to jointly optimize the model used for extracting discriminative embeddings. This SR-ASV system was evaluated on the newly released ASVspoof 2017 Version 2.0 and 2019 corpora with experimental protocols of logical and physical access scenarios. By comparing with the other state-of-the-art systems on varying types of attacks, it was proved that the SR-ASV system can offer impressive performance for the spoofing detection and speaker verification tasks. However, for some powerful spoofing techniques derived from generative models and neural networks, the performance still needs to be improved further. Future works will focus on increasing the representation ability of the model. We will also explore quantizing the proposed model to create a ``lite" version for the potential deployment in practical scenarios.  

\section*{Acknowledgment}

We gratefully acknowledge the support of NVIDIA Corporation with the donation of the Quadro P6000 GPU used for this research.

\bibliographystyle{IEEEtran}
\bibliography{IEEEabrv,mybib}

\begin{IEEEbiography}{Yuanjun Zhao}
Biography text here.
\end{IEEEbiography}

\begin{IEEEbiography}{Roberto Togneri}
	Biography text here.
\end{IEEEbiography}

\begin{IEEEbiography}{Victor Sreeram}
	Biography text here.
\end{IEEEbiography}

\end{document}